\documentclass[12pt,preprint]{aastex}

\slugcomment{RESCEU-23/01 UTAP-404}
\shorttitle{TIME DELAY STATISTICS}
\shortauthors{OGURI ET AL.}
\begin{document}
\title{Strong Gravitational Lensing Time Delay Statistics \\
and the Density Profile of Dark Halos}
%
\author{Masamune Oguri,\altaffilmark{1} Atsushi Taruya,\altaffilmark{1,2} 
Yasushi Suto,\altaffilmark{1,2} and Edwin L. Turner\altaffilmark{3}}
\email{oguri@utap.phys.s.u-tokyo.ac.jp,
ataruya@utap.phys.s.u-tokyo.ac.jp,
suto@phys.s.u-tokyo.ac.jp,
elt@astro.princeton.edu }
\altaffiltext{1}{Department of Physics, School of Science, University of
Tokyo, Tokyo 113-0033, Japan} 
\altaffiltext{2}{Research Center for the Early Universe (RESCEU), School
of Science, University of Tokyo, Tokyo 113-0033, Japan}
\altaffiltext{3}{Princeton University Observatory, Peyton Hall,
Princeton, NJ 08544, USA}
%
\received{2001 October 8}
\accepted{2001 December 5}
\begin{abstract}
The distribution of differential time delays $\Delta t$ between images
produced by strong gravitational lensing contains information on the 
mass distributions in the lensing objects as well as on cosmological
parameters such as $H_0$. We derive an explicit expression for the
conditional probability distribution function of time delays 
$P(\Delta t\,|\,\theta)$, given an image separation between multiple images
$\theta$, and related statistics. We consider lensing halos
described by the singular isothermal sphere (SIS) 
approximation and by its generalization
as proposed by Navarro, Frenk, \& White (NFW) which has a
density profile $\rho \propto r^{-\alpha}$ in the innermost
region. The time delay distribution is very sensitive to these profiles;
steeper inner slopes tend to produce larger time delays. For example,
if $H_0=70\,{\rm km\,s^{-1}Mpc^{-1}}$, a $\Lambda$-dominated cosmology and
a source redshift $z_{\rm S}=1.27$
are assumed, lenses with $\theta=5^{''}$ produce a time delay of
$\Delta t[{\rm yr}]=1.5^{+1.7}_{-0.9}$, $0.39^{+0.37}_{-0.22}$, 
$0.15^{+0.11}_{-0.09}$, and $0.071^{+0.054}_{-0.038}$ (50\% confidence 
interval),
for SIS, generalized NFW with $\alpha=1.5$, $\alpha=1.0$, and $\alpha=0.5$,
respectively. At a fixed image separation, the time delay is determined
by the difference in the lensing potential between the position of the
two images, which typically occur at different impact parameters.
Although the values of $\Delta t$ are proportional to the inverse of
$H_0$, $P(\Delta t\,|\,\theta)$ is rather insensitive to all other
cosmological model parameters, source redshifts, magnification biases
and so on. A knowledge of $P(\Delta t\,|\,\theta)$ will also be useful
in designing the observing program of future large scale synoptic
variability surveys and for evaluating possible selection biases
operating against large splitting lens systems.
\end{abstract} 
%
\section{Introduction}

The cold dark matter (CDM) scenario predicts relatively cuspy dark matter halos.
On the basis of systematic N-body simulations, \citet[hereafter
NFW]{navarro96,navarro97} found that the density profile obeys the
``universal'' form $\rho(r)\propto r^{-1}(r+r_{\rm s})^{-2}$ irrespective
of the underlying cosmological parameters, the shape of the primordial
fluctuation spectrum and the formation histories. Recent high resolution
simulations suggest even steeper cusps $\rho\propto r^{-1.5}$ in the
innermost region \citep{moore99,fukushige01a,fukushige01b}, and a
weak dependence of the inner slope on the halo mass is also reported
\citep{jing00a}.

The statistics of strong
gravitational lensing have been used
to probe the density profiles of dark halos, e.g., multiple QSO images
\citep*{fox01,keeton01,wyithe01,li01,takahashi01} and the long thin arcs
\citep*{williams99,meneghetti01,molikawa01,oguri01}. These theoretical
studies concluded that gravitational lensing rates are
extremely sensitive to the inner slope of dark halos. The lensing
statistics of small separations, however, will also be affected by
gas cooling and clumpiness in the host halo
\citep{keeton98,porciani00,kochanek01,li01}. Thus multiple QSO images
with intermediate or large separations, $\theta\gtrsim 5^{''}$, are 
more relevant in constraining the density profile of ``pure" dark halos. 

At present, several lensing surveys at large separations have been
carried out; e.g., the Jodrell-Bank VLA Astrometric Survey and the
Cosmic Lens All Sky Survey \citep[JVAS/CLASS; 
e.g.,][]{browne01} and Arcminute Radio Cluster-lens Search \citep*[ARCS;
e.g.,][]{phillips01}. The JVAS/CLASS sample comprises 10,499 radio 
sources. An explicit search for lenses has detected no lenses with image
separations $6^{''}<\theta<15^{''}$ \citep{phillips00}, while 18
gravitational lenses with $0.3^{''}<\theta<3^{''}$ were found
\citep{helbig00}. ARCS also produced a null result for lensing events with
$15^{''}<\theta<60^{''}$ from 1,023 extended radio sources.  
The lack of large separation images is only marginally consistent if the
usual NFW profile and $\Lambda$-dominated CDM model are assumed
\citep{li01,keeton01}, but it also may be ascribed in part to an 
effect of the longer time delays between more widely separated images.
Any intrinsic variability of the QSO will result in images which less
resemble each other for longer delays \citep{phillips01}.

So far, time delays between two images have been used primarily to estimate
the Hubble constant $H_0$ using a detailed lens model of each system
\citep[e.g.,][]{grogin96,barkana99}. But the importance of mass
distribution in estimating Hubble constant has been recognized 
\citep{impey98,keeton00,witt00} and this led to an attempt to constrain
the galaxy mass profile from time delays using Monte Carlo simulations
\citep{rusin00}. 
Instead, in this paper, we consider the statistics of the time delay
effect analytically. We derive an expression for
the cumulative joint probability, i.e., the probability that the
time delay is larger than $\Delta t$ and the image separation is 
$\theta$, $P(>\!\Delta t, \theta)$, and also various related statistics.
They allow us to estimate the range of probable time delays for a given
image separation and the extent to which the intrinsic time variability
of quasars affects observed strong gravitational lensing rates. 
Our most important result is that the distribution of delays is quite
sensitive to the density profiles of the lensing objects.
For example, an NFW density profile predicts median
time delays a factor of three or so smaller than the density
profile proposed by \citet{moore99} and
\citet{fukushige01a,fukushige01b} for the same $H_0$ value. While the
lenses for which time delays are currently observed are dominated by barionic
component, we can constrain the density profile of dark halos if a sample
of time delays with large separations becomes available.
It turns out that the time
delay statistics are fairly insensitive to other uncertainties, such as
the magnification bias and various cosmological parameters and therefore
become a relatively reliable estimator for the density profile of dark halos.

Of course, all delay values are linearly proportional to the inverse Hubble
constant, and we here assume its value to be $H_0=70\,{\rm
km\,s^{-1}Mpc^{-1}}$, 
which is consistent with the final result of Hubble Space
Telescope Key Project \citep{freedman01}, throughout the remainder of
our discussion.
 
The outline of this paper is as follows. In \S 2, we briefly describe
the usual formulation of gravitational lensing statistics. Section 3 presents
the analytic formulation of time delay statistics. Our main results are
shown in \S 4. Finally we summarize the main results and discuss their
application in \S 5.

\section{Description of Gravitational Lensing Statistics}
 
\subsection{Basic Equations}

We denote the image position in the lens plane by $\vec{\xi}$ and the
source position in the source plane by $\vec{\eta}$. We assume spherically
symmetric lens objects throughout the paper. In this case, the lens equation
\citep*[e.g.,][]{schneider92} reduces to 
\begin{equation}
y=x-\beta(x),
\label{lenseq}
\end{equation}
where $x=|\vec{\xi}|/\xi_0$, $y=|\vec{\eta}|D_{\rm OL}/\xi_0D_{\rm OS}$,
$\xi_0$ is the characteristic length in the lens plane (see \S 2.3.), 
and $D_{\rm OL}$ and $D_{\rm OS}$ denote the angular
diameter distances from the observer to the lens and the source planes,
respectively. The explicit expressions of the scaled deflection angles
$\beta(x)$ for the specific lens models are presented in \S 2.3.

We consider a halo of mass $M$ in the lens plane at a redshift $z_{\rm
L}$ and a source located at $z_{\rm S}$. Then the gravitational lensing
cross section $\sigma(>\!\theta, >\!\mu)$, defined in the lens plane,
for multiple images with image separation larger than $\theta$ and
magnification larger than $\mu$ is given by
\begin{equation}
\sigma(>\!\theta, >\!\mu)=\pi\,y_{\rm r}^2\,\xi_0^2\,\Theta(\theta(M, z_{\rm S}, z_{\rm L})-\theta)\,p(>\!\mu),
\end{equation}
where $y_{\rm r}$ is the critical source position to form multiple
images (usually given by the position of the radial caustic),
$\Theta(x)$ is the Heaviside step function, and $p(>\!\mu)$ denotes the
fraction of area satisfying $>\!\mu$: 
\begin{equation}
p(>\!\mu)=\frac{2}{y_{\rm r}^2}\int_0^{y_{\rm r}}dy\,y\,\Theta(\mu(y)-\mu).
\end{equation}

 From this cross section, the probability that a source at $z_{\rm S}$ is
 multiply lensed with image separation larger than $\theta$ and
 magnification larger than $\mu$ is
\begin{equation}
P(>\!\theta, >\!\mu; z_{\rm S})
=\int_0^{z_{\rm S}}dz_{\rm L}\int_{M_{\rm min}}^\infty dM\,\pi\,y_{\rm r}^2\,\xi_0^2\,p(>\!\mu)\frac{c\,dt}{dz_{\rm L}}(1+z_{\rm L})^3n(M,z_{\rm L}),
\end{equation}
where $M_{\rm min}$ is determined by solving the equation;
$\theta=\theta(M_{\rm min}, z_{\rm S}, z_{\rm L})$. 

To calculate the lens mass distribution, we employ the Press-Schechter
function \citep{press74}:
\begin{equation}
n_{\rm PS}(M, z)=\sqrt{\frac{2}{\pi}}\frac{\rho_0}{M}\frac{\delta_0(z)}{\sigma_M^2}\left|\frac{d\sigma_M}{dM}\right|\exp\left[-\frac{\delta_0^2(z)}{2\sigma_M^2}\right],
\label{ps}
\end{equation}
where $\sigma_M$ is the rms of linear density fluctuation on mass scale
$M$ at $z=0$ and $\delta_0(z)$ is the critical linear density contrast,
$\delta_0(z)\sim 1.69/D(z)$, 
with $D(z)$ being the linear growth rate normalized to unity at $z=0$.

The corresponding differential probability with respect to the image
separation is 
\begin{eqnarray}
P(\theta, >\!\mu; z_{\rm S})&\equiv&-\frac{d}{d\theta}P(>\!\theta, >\!\mu; z_{\rm S})\nonumber\\
&=&\int_0^{z_{\rm S}}dz_{\rm L}\,\frac{c\,dt}{dz_{\rm L}}(1+z_{\rm L})^3\left[\frac{dM}{d\theta}\pi\,y_{\rm r}^2\,\xi_0^2\,p(>\!\mu)n(M,z_{\rm L})\right]_{M=M_{\rm min}}.
\label{pd}
\end{eqnarray}

\subsection{Magnification Bias}

So far, we have not considered magnification bias 
\citep{turner80,turner84}. For a source of luminosity $L$, the
effect of the magnification bias may be included as follows:
\begin{eqnarray}
P^{\rm B}(\theta; z_{\rm S}, L)&=&\frac{1}{\Phi(z_{\rm S}, L)}\int_1^\infty d\mu\left|\frac{d}{d\mu}P(\theta, >\!\mu; z_{\rm S})\right|\Phi(z_{\rm S}, L/\mu)\frac{1}{\mu}\nonumber\\
&=&\int_0^{z_{\rm S}}dz_{\rm L}\,\frac{c\,dt}{dz_{\rm L}}(1+z_{\rm L})^3\left[\frac{dM}{d\theta}\pi\,y_{\rm r}^2\,\xi_0^2\,B(z_{\rm S}, L)n(M,z_{\rm L})\right]_{M=M_{\rm min}},
\label{pd_bias}
\end{eqnarray}
where $\Phi(z_{\rm S}, L)$ is the luminosity function of sources and
$B(z_{\rm S}, L)$ is
\begin{equation}
B(z_{\rm S}, L)
=\frac{2}{y_{\rm r}^2\Phi(z_{\rm S}, L)}\int_0^{y_{\rm r}}dy\,y\,\Phi(z_{\rm S}, L/\mu(y))\frac{1}{\mu(y)}.
\label{biasfactor}
\end{equation}
Thus the probability
distribution taking account of the magnification bias is simply
expressed as equation (\ref{pd}) with $p(>\!\mu)$ replaced by $B(z_{\rm
S}, L)$. By neglecting magnification bias for simplicity, we may simply
set $B(z_{\rm S}, L)=1$.

It should be noted that $\mu(y)$ may be interpreted as the magnification
of the total images, of the brighter, or of the fainter image among the
outer two images, depending on the observational selection procedure for
finding gravitational lens systems \citep{sasaki93,cen94}. As indicated
in \S 2.3.2, however, such a different choice of magnification does not
affect the conditional time delay probability very much, especially for
small inner slope values, while it changes
the lensing rate by a factor of two. 
Thus we choose to designate the magnification $\mu(y)$ in the bias factor (eq. 
[\ref{biasfactor}]) as the total magnification of all images throughout
this paper.

\subsection{Specific Density Profiles}

\subsubsection{Singular Isothermal Sphere}

The SIS (Singular Isothermal Sphere) density profile is usually 
characterized by a one-dimensional velocity dispersion $v$:
\begin{equation}
 \rho(r)=\frac{v^2}{2\pi Gr^2}.
\end{equation}
In this case, we choose the characteristic length $\xi_0$ as
\begin{equation}
 \xi_0=4\pi\left(\frac{v}{c}\right)^2\frac{D_{\rm OL}D_{\rm LS}}{D_{\rm OS}}.
\end{equation}
Then the lens equation has two solutions $x_\pm=y\pm 1$ if $|y|\leq
y_{\rm r}=1$. 
The separation between two image may be written as
\begin{equation}
 \theta=\frac{\xi_0(x_+-x_-)}{D_{\rm OL}}=8\pi\left(\frac{v}{c}\right)^2\frac{D_{\rm LS}}{D_{\rm OS}}.
\end{equation}
The magnification of each image is 
\begin{equation}
 \mu_\pm(y)=\left|\frac{y}{x_\pm}\frac{dy}{dx_\pm}\right|^{-1}=\pm\frac{x_\pm}{y}=\frac{1}{y}\pm 1,
\label{mupm_sis}
\end{equation}
and their total magnification is given by
\begin{equation}
 \mu(y)=\frac{2}{y}.
\label{muy_sis}
\end{equation}
To compute the probability distribution functions (\ref{pd}) and
(\ref{pd_bias}), 
we also convert the mass function (eq. [\ref{ps}]) to a velocity function
by using the spherical collapse model \citep[e.g.,][]{nakamura97}.

\subsubsection{Generalized NFW Profile}

The halo density profiles predicted by recent N-body simulations may be
parameterized as a one-parameter family, the generalized NFW profile
\citep{jing00a}:  
\begin{equation}
 \rho(r)=\frac{\rho_{\rm crit}\delta_{\rm c}}
{\left(r/r_{\rm s}\right)^\alpha\left(1+r/r_{\rm s}\right)^{3-\alpha}},
\label{nfw}
\end{equation}
where $r_{\rm s}$ is a scale radius and $\delta_{\rm c}$ is a
characteristic density. The profile with $\alpha=1$ corresponds that NFW
originally proposed, while the profile with $\alpha=1.5$ resembles the
one claimed by \citet{moore99} and \citet{fukushige01a,fukushige01b}. 
The scale radius generally depends on the mass and the redshift, and is
related to the concentration parameter:
\begin{equation}
c_{\rm vir}(M, z)\equiv \frac{r_{\rm vir}(M, z)}{r_{\rm s}(M, z)}.
\label{concentration}
\end{equation}
The characteristic density is given by
\begin{equation}
\delta_{\rm c}=\frac{\Delta_{\rm vir}\Omega_{\rm vir}}{3}
\frac{(3-\alpha)c_{\rm vir}^\alpha}{{}_2F_1\left(3-\alpha, 3-\alpha; 4-\alpha; -c_{\rm vir}\right)},
\end{equation}
with ${}_2F_1\left(a, b; c; x\right)$ being the hypergeometric function 
\citep[e.g.,][]{keeton01}. Following \citet{oguri01}, we adopt the
mass and redshift dependence reported by \citet{bullock01} and consider
the median amplitude of the concentration parameter $c_{\rm norm}$:
\begin{equation}
 c_{\rm vir}(M, z)=c_{\rm norm}\frac{2-\alpha}{1+z}\left(\frac{M}{10^{14}h^{-1}M_{\odot}}\right)^{-0.13},
\end{equation}
where $h$ denotes the Hubble constant in units of 
$100\,{\rm km\,s^{-1}Mpc^{-1}}$.
Scatter of the concentration parameter is modeled by a log-normal 
function with the dispersion of $\sigma_c=0.18$
\citep{jing00b,bullock01}. We fix $c_{\rm norm}$ to the value estimated 
in the simulations \citep{bullock01}, $c_{\rm norm}=8$, throughout this
paper.

In these profiles, we choose $\xi_0=r_{\rm s}$ and denote the scaled
deflection angle $\beta(x)$ as $\beta(x)=b\,f(x)$. The dimensionless
factor $b$ (of order unity) and the function $f(x)$ are related to the
dark halo profile as follows: 
\begin{equation}
b=\frac{16\pi G\rho_{\rm crit}\delta_{\rm c}r_{\rm s}}{c^2}\frac{D_{\rm OL}D_{\rm LS}}{D_{\rm OS}}, \label{b}\\
\end{equation}
\begin{equation}
f(x)=\frac{1}{x}\int_0^\infty dz \int_0^x dx' 
\frac{x'}{\left(\sqrt{x^{'2}+z^2}\right)^\alpha
\left(1+\sqrt{x^{'2}+z^2}\right)^{3-\alpha}}.
\end{equation}

The lens equation has three solutions $x_1, x_2, x_3\;(x_1>x_2>x_3)$ if
$|y|<y_{\rm r}$, where $y_{\rm r}$ is the position of the radial caustic.
The image separation is defined between the outer two solution and is
approximated as \citep{hinshaw87}:
\begin{equation}
 \theta\equiv\frac{\xi_0(x_1-x_3)}{D_{\rm OL}}\simeq\frac{2r_{\rm s}x_{\rm t}}{D_{\rm OL}},
\label{approxsep}
\end{equation}
where $x_{\rm t}$ is the position of the tangential critical curve (the
Einstein radius). The approximation (eq. [\ref{approxsep}]) is
sufficiently accurate for the range of interest here (Upper panels of Figure
\ref{fig:approx}), and is useful in computing the lensing probability. 

\begin{figure}
\plotone{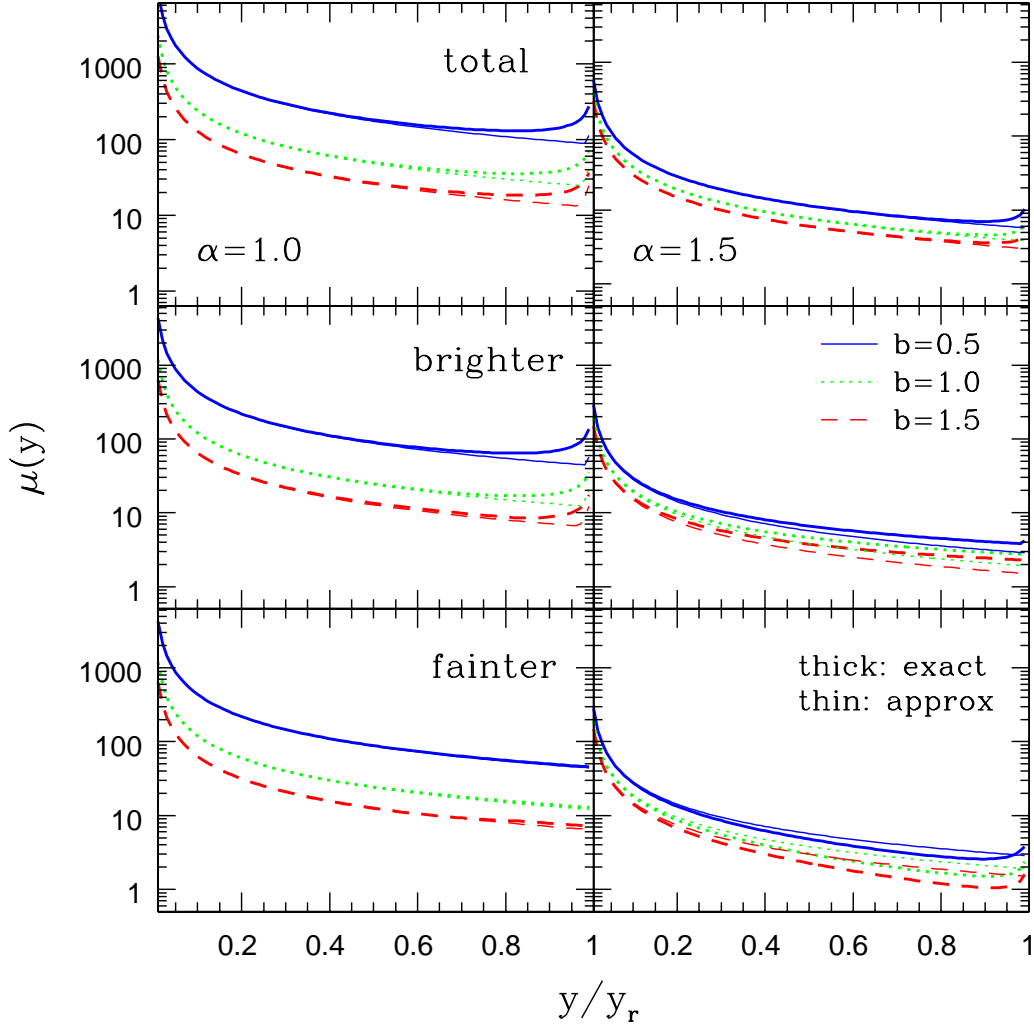} 
\caption{The magnification and its approximation defined in \S 2.3.2 
for generalized NFW profiles with various parameters $b$ and the inner
 slope $\alpha$. From top to bottom, the magnification is defined by
 that of the total images, of the brighter and of the fainter image
 among the outer two images, respectively.}  
\label{fig:approxmag}
\end{figure}

 The top panels of Figure \ref{fig:approxmag} shows that the
 total magnification for the generalized NFW profile is well approximated as
 \citep{blandford86,kovner87,nakamura96}
\begin{eqnarray}
 \mu(y)&\simeq&
\left\{
\begin{array}{@{\hspace{0.6mm}}ll}
\displaystyle{\mu_{{\rm t}0}\frac{y_{\rm r}}{y}} & \mbox{$(y < y_{\rm crit})$},\\
\displaystyle{\mu_{{\rm r}0}\frac{1}{(1-y/y_{\rm r})^{1/2}}} & \mbox{$(y > y_{\rm crit})$},
\end{array}
\right.
\label{approxmu}
\end{eqnarray}
where $\mu_{{\rm t}0}$ and $\mu_{{\rm r}0}$ are
\begin{eqnarray}
 \mu_{{\rm t}0}&=&\frac{2x_{\rm t}}{y_{\rm r}(1-bf'(x_{\rm t}))},\\
 \mu_{{\rm r}0}&=&\frac{x_{\rm r}}{y_{\rm r}}\sqrt{\frac{2}{y_{\rm r}bf''(x_{\rm r})}},
\end{eqnarray}
with $x_{\rm r}$ being the position of the radial critical curve.
Finally $y_{\rm crit}$ is given by
\begin{equation}
 \frac{y_{\rm crit}}{y_{\rm r}}=\frac{-\mu_{{\rm t}0}^2+\sqrt{\mu_{{\rm t}0}^4+4\mu_{{\rm t}0}^2\mu_{{\rm r}0}^2}}{2\mu_{{\rm r}0}^2}.
\end{equation}

The magnification of the brighter or fainter image
can be approximated as follows:
\begin{eqnarray}
 \mu_{\rm bright}(y)&\simeq&
\left\{
\begin{array}{@{\hspace{0.6mm}}ll}
\displaystyle{\frac{\mu_{{\rm t}0}}{2}\frac{y_{\rm r}}{y}} & \mbox{$(y < y_{\rm crit})$},\\
\displaystyle{\frac{\mu_{{\rm r}0}}{2}\frac{1}{(1-y/y_{\rm r})^{1/2}}} & \mbox{$(y > y_{\rm crit})$},
\end{array}
\right.
\label{approxbright}
\end{eqnarray}
\begin{equation}
 \mu_{\rm faint}(y)\simeq\frac{\mu_{{\rm t}0}}{2}\frac{y_{\rm r}}{y}.
\label{approxfaint}
\end{equation}

Figure \ref{fig:approxmag} shows the comparison between the
magnifications of the three cases and their approximation described above
(eqs. [\ref{approxmu}], [\ref{approxbright}], and [\ref{approxfaint}]).
Figure \ref{fig:approxmag} indicates that the above approximation
breaks down around its minimum value. In practice, however, this level
of discrepancy does not affect the result of the magnification bias
\citep{nakamura96}. More importantly, the difference in magnification
using total, brighter, or fainter images is only about a factor
of two, as easily seen from the above approximations. That
is, the different choices of the magnification only increase (or
decrease) the probability by a factor of two for all separations. This
numerical factor is canceled out in calculating the conditional
probability of time delays because it is defined by the ratio of these
probabilities (see \S 3.1). In the SIS case, however, the difference
between the magnification of brighter and fainter image is not
negligible (see eq. [\ref{mupm_sis}]) and as a result the effect of
different choice of magnification on time delay probability
distributions becomes larger. Therefore, we conclude that the effect of
different definitions of the magnification is small in time delay
statistics, especially for a small inner slope ($\alpha\lesssim1.5$). 
 
\section{Analytic Formulation of the Time Delay Probability Distribution}

\subsection{General Formulation}

The lens alters the time taken for light to reach the observers and
inevitably produces a differential time delay between multiple images
\citep{refsdal64,refsdal66}. The value of the time delay is calculated
as \citep[e.g.,][]{schneider92} 
\begin{equation}
 c\Delta t(y)=\frac{\xi_0^2D_{\rm OS}}{D_{\rm OL}D_{\rm LS}}(1+z_{\rm L})\left[\phi(x^{(1)}, y)-\phi(x^{(2)}, y)\right],
\label{timedelay}
\end{equation}
where $x^{(i)}$ ($i=1,2$) are two image positions. The Fermat potential
is written as 
\begin{equation}
\phi(x, y)=\frac{(x-y)^2}{2}-\psi(x),
\end{equation}
in terms of the lensing potential $\psi(x)$. The explicit expressions
for the time delays in the SIS and the generalized NFW profiles are
given in \S 3.2. 

The cumulative probability of time delays is calculated from equations
(\ref{pd_bias}), (\ref{biasfactor}), and (\ref{timedelay}): 
\begin{equation}
 P^{\rm B}(>\!\Delta t, \theta; z_{\rm S}, L)=\int_0^{z_{\rm S}}dz_{\rm L}\,\frac{c\,dt}{dz_{\rm L}}(1+z_{\rm L})^3\left[\frac{dM}{d\theta}\pi\,y_{\rm r}^2\,\xi_0^2\,B^{\rm T}(>\!\Delta t; z_{\rm S}, L)n(M,z_{\rm L})\right]_{M=M_{\rm min}},
\label{delay_pd}
\end{equation}
\begin{eqnarray}
 B^{\rm T}(>\!\Delta t; z_{\rm S}, L)=
\left\{
\begin{array}{@{\hspace{0.6mm}}ll}
\displaystyle{\frac{2}{y_{\rm r}^2\Phi(z_{\rm S}, L)}\int_{y_{\rm min}}^{y_{\rm r}}dy\,y\,\Phi(z_{\rm S}, L/\mu(y))\frac{1}{\mu(y)}} & \mbox{($y_{\rm min}<y_{\rm r}$)}, \\
\displaystyle{0} & \mbox{($y_{\rm min}>y_{\rm r}$)},
\end{array}
\right.
\label{delay_biasfactor}
\end{eqnarray}
where the superscript $T$ on the magnification factor means that the time
delay threshold is taken into account. The lower limit of integral
$y_{\rm min}$ is determined by solving the equation; $\Delta t=\Delta
t(y_{\rm min})$. This expression is valid when the time delay is a
monotonic function of the source position $y$ as in all our examples below. 
 
When the magnification bias is neglected, we can replace $B^{\rm
T}(>\!\Delta t; z_{\rm S}, L)$ by
\begin{eqnarray}
 N^{\rm T}(>\!\Delta t)=
\left\{
\begin{array}{@{\hspace{0.6mm}}ll}
\displaystyle{\frac{2}{y_{\rm r}^2}\int_{y_{\rm min}}^{y_{\rm r}}dy\,y=1-\left(\frac{y_{\rm min}}{y_{\rm r}}\right)^2} & \mbox{($y_{\rm min}<y_{\rm r}$)}, \\
\displaystyle{0} & \mbox{($y_{\rm min}>y_{\rm r}$)}.
\end{array}
\right.
\end{eqnarray}

The cumulative conditional probability can be calculated from
equations (\ref{pd}) and (\ref{delay_pd}):
\begin{equation}
 P^{\rm B}(>\!\Delta t\,|\,\theta; z_{\rm S}, L)=\frac{P^{\rm B}(>\!\Delta t, \theta; z_{\rm S}, L)}{P^{\rm B}(\theta; z_{\rm S}, L)}.
\label{cond_p}
\end{equation}

We define the joint probability distribution of time delays as: 
\begin{eqnarray}
P^{\rm B}(\Delta t, \theta; z_{\rm S}, L)&\equiv&-\frac{d}{d(\Delta t)}P^{\rm B}(>\!\Delta t, \theta; z_{\rm S}, L)\nonumber\\
&=&\int_0^{z_{\rm S}}dz_{\rm L}\,\frac{c\,dt}{dz_{\rm L}}(1+z_{\rm L})^3\left[\frac{dM}{d\theta}\pi\,y_{\rm r}^2\,\xi_0^2\,B^{\rm T}(\Delta t; z_{\rm S}, L)n(M,z_{\rm L})\right]_{M=M_{\rm min}},
\end{eqnarray}
\begin{eqnarray}
 B^{\rm T}(\Delta t; z_{\rm S}, L)
&=&
\left\{
\begin{array}{@{\hspace{0.6mm}}ll}
\displaystyle{\frac{2}{y_{\rm r}^2\Phi(z_{\rm S}, L)}\left[\frac{dy}{d(\Delta t)}y\,\Phi(z_{\rm S}, L/\mu(y))\frac{1}{\mu(y)}\right]_{y=y_{\rm min}}} & \mbox{($y_{\rm min}<y_{\rm r}$)}, \\
\displaystyle{0} & \mbox{($y_{\rm min}>y_{\rm r}$)}.
\end{array}
\right.
\label{td_biasfactor}
\end{eqnarray}
Similarly, when the magnification bias is not taken into account,
we replace $B^{\rm T}(\Delta t; z_{\rm S}, L)$ by
\begin{eqnarray}
 N^{\rm T}(\Delta t)=
\left\{
\begin{array}{@{\hspace{0.6mm}}ll}
\displaystyle{\frac{2}{y_{\rm r}^2}\left[\frac{dy}{d(\Delta t)}y\right]_{y=y_{\rm min}}} & \mbox{($y_{\rm min}<y_{\rm r}$)}, \\
\displaystyle{0} & \mbox{($y_{\rm min}>y_{\rm r}$)}.
\end{array}
\right.\label{nt}
\end{eqnarray}

Finally the conditional probability distribution of time delays is
calculated as in equation (\ref{cond_p}):
\begin{equation}
 P^{\rm B}(\Delta t\,|\,\theta; z_{\rm S}, L)=\frac{P^{\rm B}(\Delta t, \theta; z_{\rm S}, L)}{P^{\rm B}(\theta; z_{\rm S}, L)}.
\label{cond_pd}
\end{equation}

\subsection{SIS and Generalized NFW Profiles}

The time delay in the SIS case is calculated from the lensing potential,
$\psi(x)=|x|$, and equation (\ref{timedelay}):
\begin{equation}
 c\Delta t(y)=32\pi^2\left(\frac{v}{c}\right)^4\frac{D_{\rm OL}D_{\rm LS}}{D_{\rm OS}}(1+z_{\rm L})y,
\label{tdelay_sis}
\end{equation}

\begin{figure}
\plotone{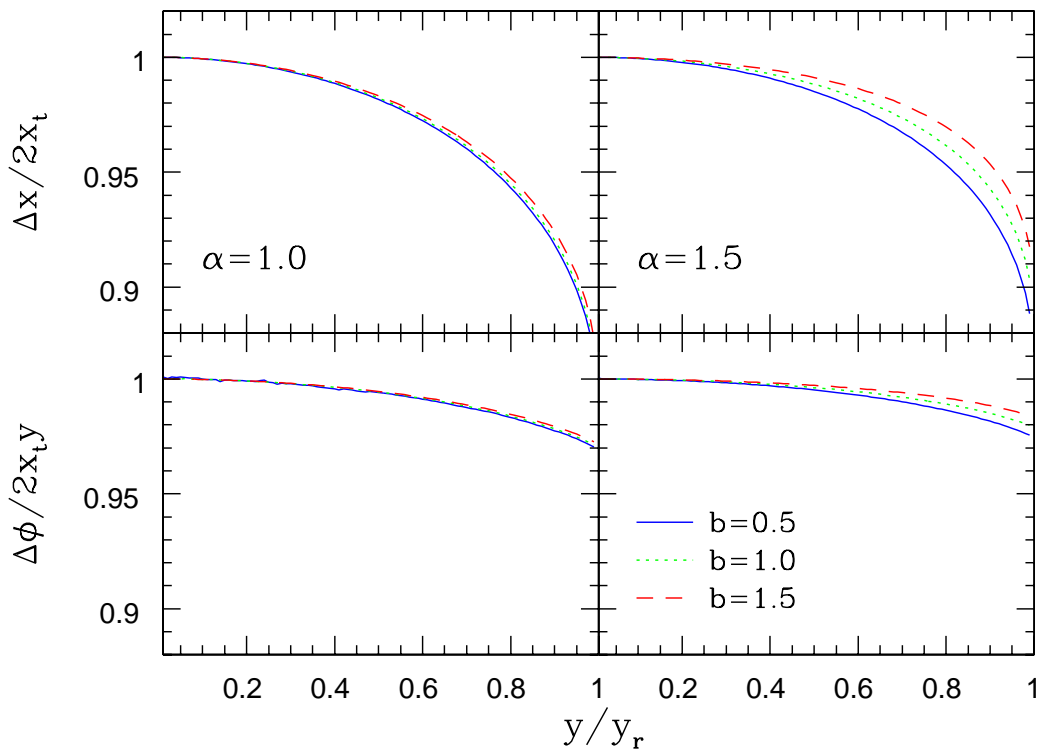} 
\caption{Accuracy of approximations for generalized
 NFW profiles with various parameters $b$ (eq. [\ref{b}]) and the inner
 slope $\alpha$; $\alpha=1.0$ ({\it left panels}) and $\alpha=1.5$ ({\it
 right panels}). {\it Upper panels} display the ratio of the image
 separation and its approximation (eq. [\ref{approxsep}]). {\it Lower
 panels} give the ratio of the difference of the Fermat potential
 and its approximation (eq. [\ref{approxphi}]).}  
\label{fig:approx}
\end{figure}

In the generalized NFW case, it is useful to adopt the following
approximation in calculating the time delay:
\begin{equation}
 \phi(x_3, y)-\phi(x_1, y)\simeq 2x_{\rm t} y.
 \label{approxphi}
\end{equation}
The accuracy of this approximation is shown in Figure \ref{fig:approx}
 ({\it Lower panels}). In the analysis below, we employ the
 approximations shown in Figures \ref{fig:approx} and
 \ref{fig:approxmag}. From equations (\ref{timedelay}) and
 (\ref{approxphi}), the time delay for the generalized NFW case is given
 by 
\begin{equation}
c\Delta t(y)=\frac{2r_{\rm s}^2x_{\rm t}D_{\rm OS}}{D_{\rm OL}D_{\rm LS}}(1+z_{\rm L})y.
\label{tdelay_nfw}
\end{equation}

\section{Results}

\subsection{Setting the Conditions}

In what follows, we consider three representative cosmological
models dominated by CDM; Lambda CDM (LCDM) with $(\Omega_0, \lambda_0, h,
\sigma_8)=(0.3, 0.7, 0.7, 1.04)$, Standard CDM (SCDM) with $(1.0, 0.0,
0.7, 0.54)$, and Open CDM (OCDM) with $(0.45, 0.0, 0.7, 0.83)$. The
amplitude of the mass fluctuation, $\sigma_8$, is normalized so as to
reproduce the X-ray luminosity and temperature functions of clusters
\citep{kitayama97}.

We calculate the probability distribution of image separations and time
delays adopting the detection condition in
the JVAS/CLASS survey. The sample of radio sources have a flux
distribution with $\Phi(S)\propto S^{-2.1}$ \citep{rusin01}. In this
case, magnification bias factors (\ref{biasfactor}) and
(\ref{delay_biasfactor}) are simplified and can be written as
\begin{equation}
B(z_{\rm S}, L)=\frac{2}{y_{\rm r}^2}\int_0^{y_{\rm r}}dy\,y\,\left\{\mu(y)\right\}^{\gamma-1},
\label{simple_biasfactor}
\end{equation}
and
\begin{eqnarray}
 B^{\rm T}(>\!\Delta t; z_{\rm S}, L)=
\left\{
\begin{array}{@{\hspace{0.6mm}}ll}
\displaystyle{\frac{2}{y_{\rm r}^2}\int_{y_{\rm min}}^{y_{\rm r}}dy\,y\,\left\{\mu(y)\right\}^{\gamma-1}} & \mbox{($y_{\rm min}<y_{\rm r}$)}, \\
\displaystyle{0} & \mbox{($y_{\rm min}>y_{\rm r}$)}.
\end{array}
\right.
\end{eqnarray}
with $\gamma$ being the slope of the radio luminosity function, $\gamma=2.1$. 
Using the fact that the luminosity function is described by a power-law
and the time delay is proportional to the source position (i.e., $\Delta
t\propto y$), equations (\ref{td_biasfactor}) and (\ref{nt}) are also
simplified as 
\begin{eqnarray}
 B^{\rm T}(\Delta t; z_{\rm S}, L)=
\left\{
\begin{array}{@{\hspace{0.6mm}}ll}
\displaystyle{\frac{2y_{\rm min}^2}{\Delta t\,y_{\rm r}^2}\left\{\mu(y_{\rm min})\right\}^{\gamma-1}} & \mbox{($y_{\rm min}<y_{\rm r}$)}, \\
\displaystyle{0} & \mbox{($y_{\rm min}>y_{\rm r}$)},
\end{array}
\right.
\label{simple_delay_biasfactor}
\end{eqnarray}
and
\begin{eqnarray}
 N^{\rm T}(\Delta t)=
\left\{
\begin{array}{@{\hspace{0.6mm}}ll}
\displaystyle{\frac{2y_{\rm min}^2}{\Delta t\,y_{\rm r}^2}} & \mbox{($y_{\rm min}<y_{\rm r}$)}, \\
\displaystyle{0} & \mbox{($y_{\rm min}>y_{\rm r}$)}.
\end{array}
\right.
\end{eqnarray}
The redshift distribution of the parent 
population of radio sources is not known in the JVAS/CLASS
survey. Thus we fix the source redshift to the mean redshift of a 27 object
subsample, $\langle z_{\rm S}\rangle=1.27$ \citep{marlow00}. The
results are not substantially different if one takes
into account the observed redshift distribution of the subsample \citep{keeton01}.

\subsection{Time Delay Probability Distribution}

\begin{figure}
\plotone{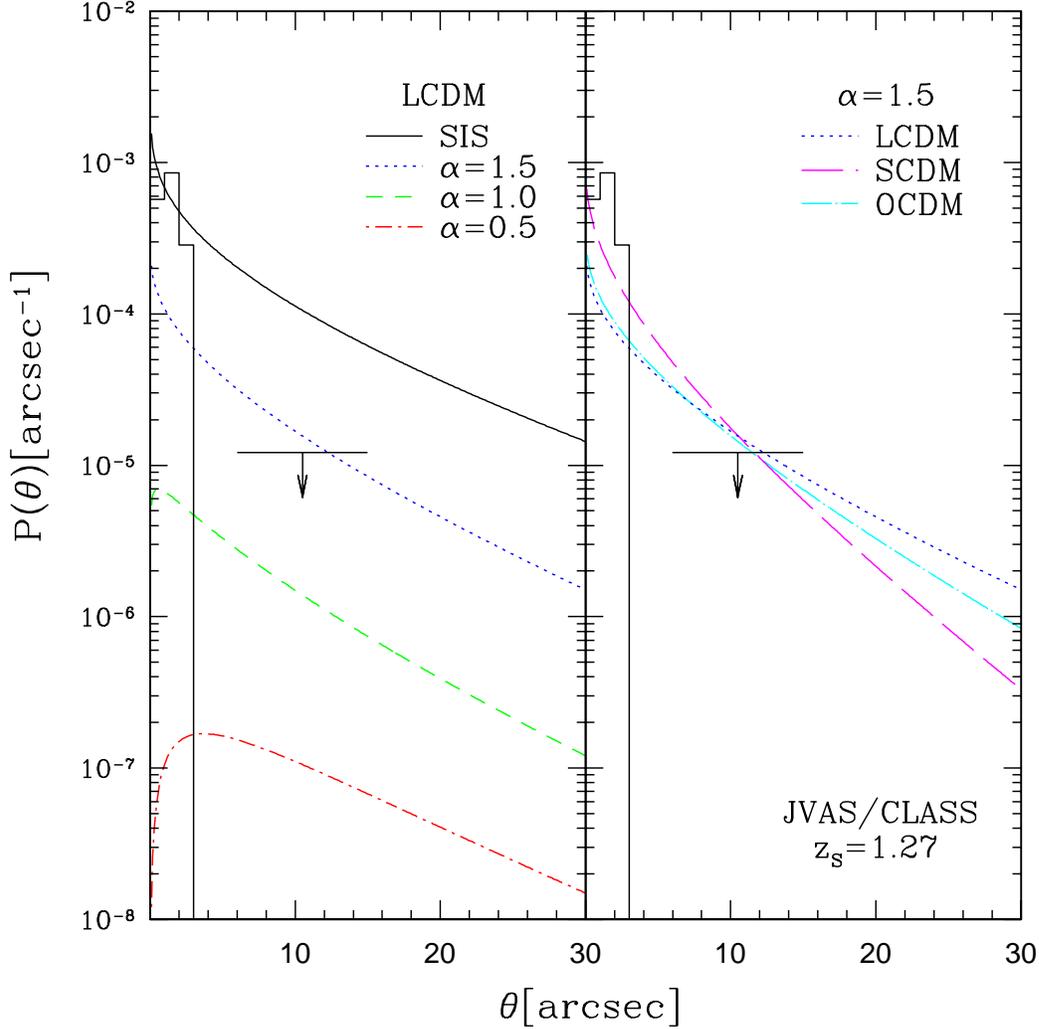} 
\caption{Differential Probability of lensing in the JVAS/CLASS survey. The
 magnification bias is included. {\it Left panel} shows  
 the probability for various density profiles; SIS ({\it solid}),
 generalized NFW with $\alpha=1.5$ ({\it dotted}), $\alpha=1.0$ ({\it
 short dashed}), and $\alpha=0.5$ ({\it short dash-dotted}). {\it Right
 panel} is the plot for various cosmological models; LCDM ({\it
 dotted}), SCDM ({\it long dashed}), and OCDM ({\it long dash-dotted}).
 The observational value is shown by the histogram. The $1\sigma$
 constraint from the JVAS/CLASS null result is also shown by a
 horizontal line with a downward arrow.}  
\label{fig:prob}
\end{figure}

Figure \ref{fig:prob} plots the predicted probability distribution
of image separations for various density profiles ({\it Left panel}); SIS,
generalized NFW with $\alpha=1.5$, $\alpha=1.0$, and $\alpha=0.5$, 
and also for various cosmological models ({\it Right panel}) fixing
$\alpha=1.5$, for reference. As shown in these plots, the probability
strongly depends on the inner profile of dark halos. The steeper inner
profile produces multiple lenses much more efficiently. The JVAS/CLASS data
are also shown in Figure \ref{fig:prob}. This plot indicates
that small separation lensing is consistent with the SIS model, while
the SIS model predicts too much large separation lensing, as previously
shown by \citet{li01}.

\begin{figure}
\plotone{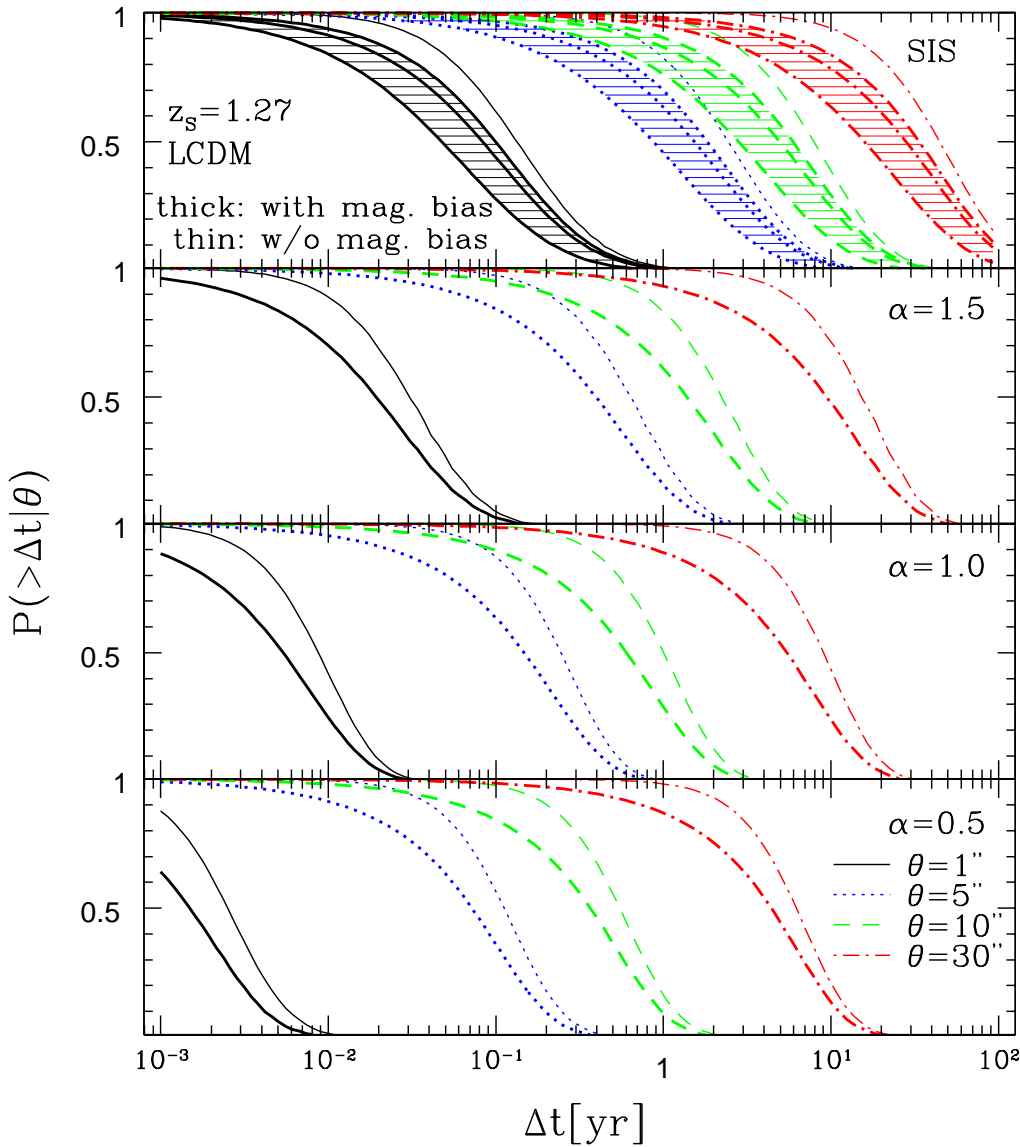} 
\caption{The cumulative conditional probability of time delays (eq.
 [\ref{cond_p}]) 
 for various density profiles with image separations $\theta=1^{''}$
 ({\it solid}), $5^{''}$ ({\it dotted}), $10^{''}$ ({\it dashed}), and
 $30^{''}$ ({\it dash-dotted}). The cosmological model is LCDM in each case. 
 Results, with ({\it thick}) and without ({\it thin}) the
 magnification bias, are presented. At the {\it top panel}, the effect of
 different choice of magnification is shown by the same three lines and
 shadings; magnification is defined by that of total images ({\it center
 lines}), of the brighter ({\it right lines}) and of the fainter image
 among the outer two images ({\it left lines}).}
\label{fig:cuml}
\end{figure}

\begin{figure}
\plotone{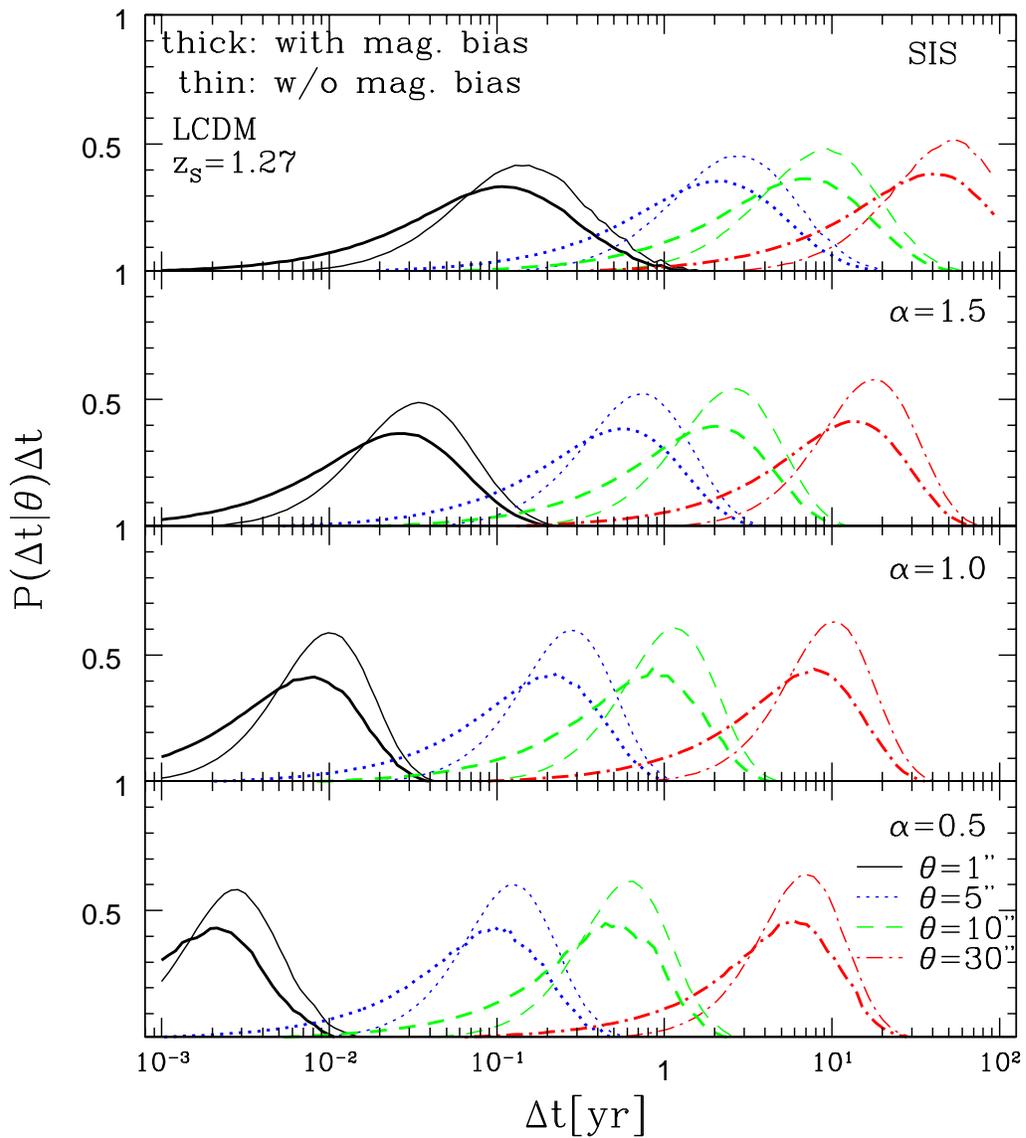}
\caption{Same as Figure \ref{fig:cuml}, but the
 differential conditional probability distribution 
 of time delays times $\Delta t$
 (eq. [\ref{log_cond_pd}]) is plotted.}
\label{fig:diff_ano}
\end{figure}

Consider next the time delay probability distributions. Figure
\ref{fig:cuml} plots the cumulative conditional probability (eq.
[\ref{cond_p}]) for different density profiles, and Figure
\ref{fig:diff_ano} plots its logarithmic time derivative:
\begin{equation}
 \frac{d}{d(\ln\Delta t)}P(>\!\Delta t\,|\,\theta; z_{\rm S}, L)=P(\Delta t\,|\, \theta; z_{\rm S}, L)\Delta t.
\label{log_cond_pd}
\end{equation}
The cosmological model is LCDM in each case. We present results both
with and without the magnification bias effects. In the Top panel of Figure
\ref{fig:cuml}, we plot the three cases of magnification bias defined by
that of the total, brighter, and fainter images. The effect of different
choice of magnification in the generalized NFW case is smaller than in the SIS
case, as discussed in \S 2.3.2. These plots indicate that time
delay statistics depend strongly on the inner slope of the density
profile, $\alpha$. Steeper inner profiles tend to produce longer time
delays. This is probably because of the deeper gravitational potential
for steeper inner profiles. The non-geometric part of the time delay is simply
proportional to the difference in the potential value at the image positions.
Therefore, we can constrain the density
profile of dark halos from this statistics if a reliable and unbiased
sample of time delays becomes available.

We note the difference in the asymptotic behavior in the limit of small
$\Delta t$ for different magnification biases.  
For the SIS profile, we obtain the asymptotic behavior of the
conditional probability distribution in the limit of the small $\Delta
t$ from equations (\ref{muy_sis}), (\ref{tdelay_sis}) and
(\ref{simple_delay_biasfactor}):
\begin{equation}
 P(\Delta t\,|\,\theta; z_{\rm S}, L)\propto(\Delta t)^{2-\gamma}.
\label{asympt}
\end{equation}
For the generalized NFW profile, we also obtain the same relation from
equations (\ref{approxmu}), (\ref{tdelay_nfw}) and
(\ref{simple_delay_biasfactor}). The asymptotic behavior (\ref{asympt})
indicates that the uncertainty in the slope of the QSO luminosity function
severely affects the probability of shorter time delays. We are not concerned
with details of $P(\Delta t\,|\,\theta)$ at small $\Delta t$,
where various uncertainties such as the finite size of the sources may
affect the shape of $P(\Delta t\,|\,\theta)$ anyway. 
Except for the small
$\Delta t$ regime, however, the uncertainty of the magnification bias is
negligible and the time delay probability distribution depends
primarily on the inner slope of lens density profile. 
In fact, the
cumulative conditional probability $P(>\!\Delta t\,|\,\theta)$ with or
without magnification bias is quite similar and probably observationally
indistinguishable for the foreseeable future.
In particular, the insensitivity of magnification bias shown
in Figures \ref{fig:cuml} and \ref{fig:diff_ano} is in marked contrast
with the usual lensing probability as a function of image separation
which is affected by more than one order of magnitude
\citep{wyithe01,li01,takahashi01}. 

To see the dependence of separations and density profiles on the
conditional probability distributions more clearly, we calculate the
median time delays $\Delta t_{\rm med}$ as follows: 
\begin{equation}
 P(>\!\Delta t_{\rm med}\,|\,\theta; z_{\rm S}, L)=\frac{1}{2}.
\label{med}
\end{equation}
We plot the $\Delta t_{\rm med}$ as a function of separations $\theta$
in Figure \ref{fig:meddis_obs}. The comparison with the observation in
this figure will be discussed in \S 4.3. The error-bars in this plot
are defined by the $\pm 25\%$ level. This figure shows that the median
time delays are well fitted by the power-law: 
\begin{equation}
 \Delta t_{\rm med}=A(\theta/1^{''})^{B}.
\label{fit}
\end{equation}
The best-fit values for $A$ and $B$ are summarized in Table
\ref{table:fit}. Since $B\sim 2$ for all density profiles, Figure 
\ref{fig:meddis_alpha_log} plots $\Delta t_{\rm med}/\theta^2$ to
illustrate their strong dependence on $\alpha$. It is clear in this
figure that the difference of $\Delta t_{\rm med}$ between profiles with
various $\alpha$ values is larger for smaller separations. To probe the density
profile of dark halos, however, small separation lenses may suffer from
the complex physics on baryonic components. Therefore lenses with
$5^{''}\lesssim\theta\lesssim10^{''}$, which are typically associated
with galaxy groups or clusters, may be more useful in constraining the
density profile of dark halos. Figure \ref{fig:meddis_alpha_log} also
indicates that it is difficult to determine the inner slope from a
single observed lensing system, and several lensing systems are required
to reduce the statistical uncertainty.

\begin{figure}
\plotone{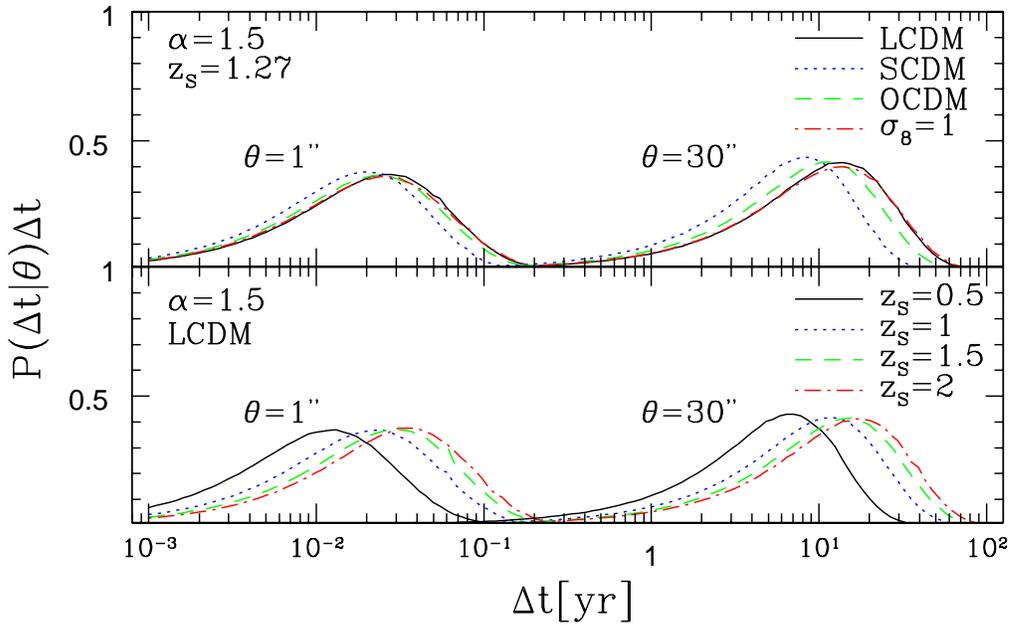} 
\caption{Dependence of 
 the conditional probability distribution of time delays on
 cosmological model and source redshift.
 A generalized NFW
 density profile with $\alpha=1.5$ is adopted. In the upper panel,
 the cosmological models are LCDM ({\it solid}) , SCDM ({\it dotted}),
 and OCDM ({\it dashed}). {\it Dash-dotted lines} are also for SCDM
 case, but in this case $\sigma_8=1$ instead of $\sigma_8=0.54$.
In the lower panel, source redshifts
 $z_{\rm S}=0.5$ ({\it solid}), $z_{\rm S}=1.0$ ({\it dotted}),  
$z_{\rm S}=1.5$ ({\it dashed}), and $z_{\rm S}=2.0$ ({\it dash-dotted})
 are plotted.}
\label{fig:diff_model}
\end{figure}

Next consider the dependence of the statistics on cosmological models or
source redshift. Figure \ref{fig:diff_model} displays the predictions
for different models and source redshifts, indicating that their effects
are rather small. In particular, we emphasize that the time delay 
probability distribution is insensitive to $\sigma_8$, while the SCDM
model with $\sigma_8=1$  yields nearly two orders more multiple lenses
compared with $\sigma_8=0.54$. The time delay probability distribution
is also insensitive to the redshift uncertainty, especially at high 
redshift $z_{\rm S}\gtrsim 1$. Therefore, uncertainties of both cosmological
models and redshift are not a significant source of uncertainty in
time delay statistics. This robustness is also an advantage of time delay
statistics compared with the usual overall lensing rate statistic.

\subsection{Comparison with Existing Observational Data}

\begin{figure}
\plotone{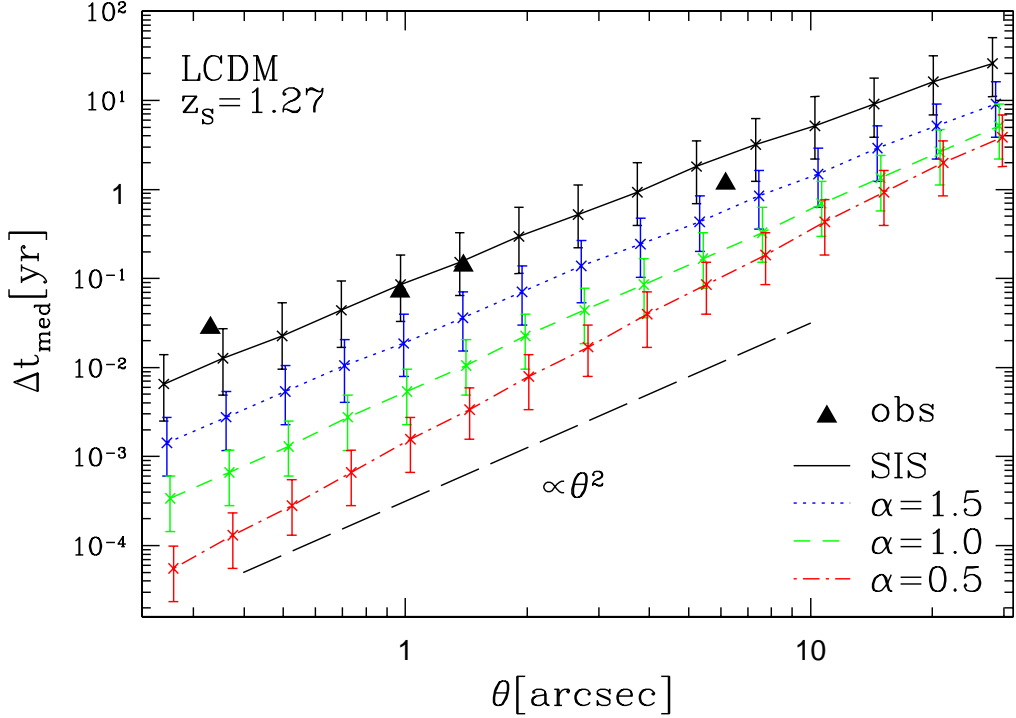} 
\caption{Median time delays, $\Delta t_{\rm med}$, as a function of
 separations $\theta$, for various density profiles. The median is
 calculated by equation (\ref{med}). The error-bars in this plot are
 defined by the $\pm 25\%$ level. A line with $\Delta t\propto\theta^2$ is
 also shown for reference. Some observational data 
 (see Table 2) are indicated by
 triangles.}
\label{fig:meddis_obs}
\end{figure}

\begin{figure}
\plotone{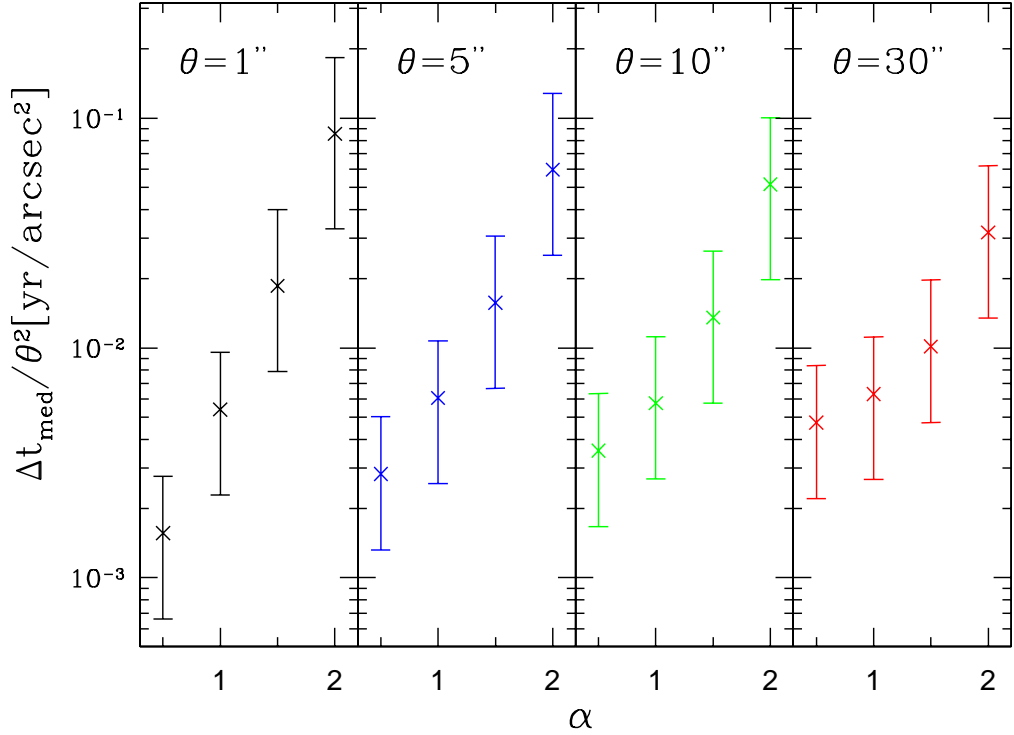} 
\caption{Dependence of time delays on inner slope $\alpha$.
 Median time delays divided by the square of the separations, 
$\Delta t_{\rm med}/\theta^2$, are plotted. In this plot, the SIS is
 regarded as $\alpha=2.0$. The error-bars are the same as in Figure
 \ref{fig:meddis_obs}.}
\label{fig:meddis_alpha_log}
\end{figure}

\begin{figure}
 \plotone{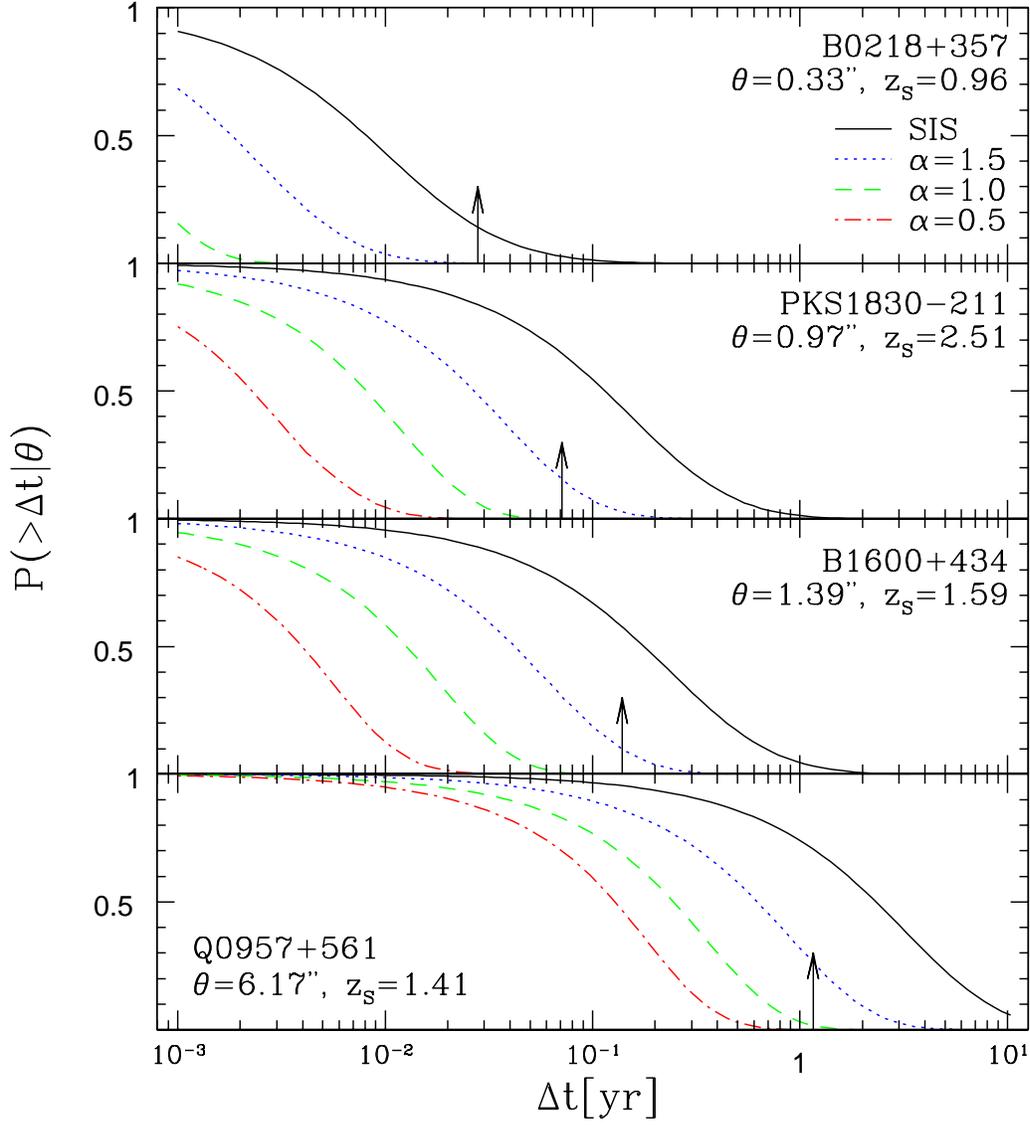} 
\caption{Comparison of predictions with existing time delay data using a
 cumulative conditional probability.  The sample of the lens systems
 used here is summarized in Table \ref{table:obs}. 
Arrows show the observed values of the time delay
 in each system. Theoretical predictions
with various density profiles are displayed by lines; SIS ({\it
 solid}), generalized NFW with $\alpha=1.5$ ({\it dotted}), $\alpha=1.0$
 ({\it dashed}), and $\alpha=0.5$ ({\it dash-dotted}). In calculating
 the theoretical distributions, the LCDM model was assumed.}
\label{fig:cuml_obs}
\end{figure}

Here we tentatively compare our theoretical results with
the existing observational data, although the detailed modeling is
already available for each system in our small sample. The adopted data are
summarized in Table \ref{table:obs}. We use time delay data for
two images systems and exclude the four image lens systems for simplicity.
The data points in Figure \ref{fig:meddis_obs} and Figure \ref{fig:cuml_obs}
show the comparison with observations for four lens systems. 
Although Figure \ref{fig:cuml_obs} displays a more direct comparison, 
Figure \ref{fig:meddis_obs} may be useful also in examining whether
there is any dependence of the inner slope $\alpha$ on the mass of the
lensing objects.
For Figure \ref{fig:meddis_obs}, we fix the source redshift at $z_{\rm
S}=1.27$, although the source redshift for each system is
already known. This does not make any substantial differences as
indicated in Figure \ref{fig:diff_model}. We plot the theoretical
predictions of SIS and generalized NFW density profiles. Observational
time delays are indicated by triangles (in Fig. \ref{fig:meddis_obs})
and arrows (in Fig. \ref{fig:cuml_obs}). From these plots, we find that
the existing data prefer the steeper inner profiles, SIS profile or
generalized NFW profile with $\alpha=1.5$. Although a different choice of
magnification affects the median time delays by at most a factor of two (see
Fig. \ref{fig:cuml}) in the SIS case and less in the
generalized NFW case, this is smaller than the error-bars of median time
delays and does not change our basic conclusion. The
steep density profiles we infer are also consistent with the detailed models
\citep[e.g.,][]{keeton00} of these particular lens systems. 
Of course, the image separations
in the lens sample considered here correspond to the inner regions of the
lensing galaxies where a plethora of evidence (especially rotation curves)
already lead us to expect a roughly isothermal distribution of the total
mass, luminous and dark combined, so the results are hardly surprising.

\section{Summary and Discussions}

In this paper, we formulated the differential time delay probability
distribution and examined its dependence on lens density profiles as
well as on magnification bias, cosmological model, and source redshift.
We found that the probability distribution of time delays depends most
sensitively on the  inner slope of the lens density profiles. The
difference between the various density  
profiles we examined is very large, more than one order. For
example, if $H_0=70\,{\rm km/s/Mpc}$, $\Lambda$-dominated cosmology and
 a source redshift $z_{\rm S}=1.27$
 are assumed, lenses with $\theta=5^{''}$ induce the time delay; 
$\Delta t[{\rm yr}]=1.5^{+1.7}_{-0.9}$, $0.39^{+0.37}_{-0.22}$, 
$0.15^{+0.11}_{-0.09}$, and $0.071^{+0.054}_{-0.038}$ (50\% level), for
SIS, generalized NFW with $\alpha=1.5$, $\alpha=1.0$, and $\alpha=0.5$,
respectively. 

On the other hand, the effects of cosmological models, source
redshifts or magnification bias are rather small, aside from the well known
and important linear dependence on the inverse Hubble constant.
Moreover, the delay distributions are
quite insensitive to the normalization of the overall lensing rate, 
because the conditional probability is defined by the joint probability
divided by the usual lensing probability and the differing of normalizations
are almost canceled out. Therefore, one could strongly constrain the core
structure of dark halos if a large sample of time delays becomes
available in future systematic survey. 

Although existing lens
systems are individually modeled in detail 
including the central density profiles, such
careful treatment of each lens system in upcoming very large samples 
may not always be practical.
Thus a {\it statistical} treatment of time delays is informative. 
For example, we can predict the range of probable time delays of
large separation lenses from this formulation if the density profile of
dark halos is fully settled. This would allow estimation of a plausible
range of time delays for {\it future} lens systems even when 
the lensing object has not been
identified.

Comparison with the meager existing observational data suggests that the
density profile has a rather steep cusp, $\alpha\sim2$, although it is
already known that the current observed systems are dominated by
baryonic component.
Moreover, it seems that the smaller separation lenses prefer steeper
inner profiles. This result may support a two-population model, i.e.,
galactic mass halos with a steep inner slope ($\alpha\sim2$) and cluster
mass halos with a shallower inner slope ($\alpha\lesssim1.5$), as
proposed by \citet{keeton98} (see also \citealp{porciani00,li01}). To
see whether  
there are two (or more) populations of halos, time delay data for various
separations, especially separations with
$5^{''}\lesssim\theta\lesssim10^{''}$, will be essential.

Another application of time delay statistics is that future
large samples of lenses will probably be systematically monitored by one
or more of the ambitious synoptic surveys now being planned. Designing an
efficient sampling rate and observing strategy requires an idea of the
range of time delays that might reasonably be expected. 
 
\acknowledgments
We thank Eiichiro Komatsu, Kentaro Nagamine, and Ryuichi Takahashi for
discussions and comments. We also thank an anonymous referee for pointing
out several important references. This research was supported in part by the
Grant-in-Aid by the Ministry of Education, Science, Sports and Culture
of Japan (07CE2002) to RESCEU and by NSF grant AST98-02802.
\clearpage

\clearpage
\begin{deluxetable}{ccc}
\tablewidth{0pt}
 \tablecaption{Fitting Parameters of Median Time Delays\label{table:fit}}
\tablehead{\colhead{Model} & \colhead{$A$[yr]} & \colhead{$B$}}
\startdata
SIS        & $8.05\times10^{-2}$ & 1.77 \\
$\alpha$=1.5 & $1.91\times10^{-2}$ & 1.86 \\
$\alpha$=1.0 & $5.37\times10^{-3}$ & 2.05 \\
$\alpha$=0.5 & $1.48\times10^{-3}$ & 2.38 \\
\enddata
\tablecomments{
Fitting Parameters $A$ and $B$ are defined in equation (\ref{fit})}
\end{deluxetable}
\begin{deluxetable}{ccccc}
\tablewidth{0pt}
\tablecaption{Observed Time Delays for Lenses with Two Images \label{table:obs}}
\tablehead{
\colhead{Lens}            & \colhead{$z_{\rm S}$}  &
\colhead{$\Delta t$ [yr]} & \colhead{$\theta$}     &
\colhead{Ref.}}
\startdata
B0218$+$357   & 0.96 & 0.028 $\pm$ 0.004 & $0.33^{''}$ & 1, 2 \\
PKS1830$-$211 & 2.51 & 0.071 $\pm$ 0.014 & $0.97^{''}$ & 3, 4 \\
B1600$+$434   & 1.59 & 0.139 $\pm$ 0.011 & $1.39^{''}$ & 5, 6 \\
Q0957$+$561   & 1.41 & 1.157 $\pm$ 0.002 & $6.17^{''}$ & 7, 8 \\
\enddata
\tablerefs{
(1) Cohen et al. 2000; (2) Biggs et al. 1999; (3) Lovell et al. 1998;
 (4) Leh{\' a}r et al. 2000; (5) Burud et al. 2000; (6) Koopmans et al.
 2000; (7) Oscoz et al. 2001; (8) Kundi{\' c} et al. 1997}
\end{deluxetable}
\clearpage
\clearpage
\clearpage
\clearpage
 \clearpage
 \clearpage
 \clearpage
\clearpage
\clearpage
\end{document}